# Cyber-Security in Smart Grid: Survey and Challenges

Z. Elmrabet[1], H. Elghazi[1], N. Kaabouch[2], H. Elghazi[1]

[1]STRS Lab, INPT, Rabat, Morocco

[2]Electrical Engineering Department, UND, USA

**Abstract--- Smart grid uses the power of information technology to intelligently deliver energy to customers by using a two-way communication, and wisely meet the environmental requirements by facilitating the integration of green technologies. Although smart grid addresses several problems of the traditional grid, it faces a number of security challenges. Because communication has been incorporated into the electrical power with its inherent weaknesses, it has exposed the system to numerous risks. Several research papers have discussed these problems. However, most of them classified attacks based on confidentiality, integrity, and availability, and they excluded attacks which compromise other security criteria such as accountability. In addition, the existed security countermeasures focus on countering some specific attacks or protecting some specific components, but there is no global approach which combines these solutions to secure the entire system. The purpose of this paper is to provide a comprehensive overview of the relevant published works. First, we review the security requirements. Then, we investigate in depth a number of important cyber-attacks in smart grid to diagnose the potential vulnerabilities along with their impact. In addition, we proposed a cyber security strategy as a solution to address breaches, counter attacks, and deploy appropriate countermeasures. Finally, we provide some future research directions.**

*Index Terms—* Smart grid, cyber-attacks, vulnerabilities, confidentiality, availability, integrity, accountability, IDS, cryptography, network security.

## I. INTRODUCTION

Traditional electrical distribution systems are used to transport electrical energy generated at a central power plant by increasing voltage levels and then delivering it to the end users by reducing voltage levels gradually. However, this electricity grid has major shortcomings, including the inability to include diverse generation sources such as green energy, high cost and expensive assets, time consuming demand response, high carbon emission, and blackouts. For example, a study conducted by researchers at the Berkeley National Laboratory in 2004 showed that power interruptions cost the American economy approximately $80 billion per year; other estimates indicate a higher cost of $150 billion per year [1]. It is evident that these critical problems cannot be addressed with existing electricity grid. Smart grid promises to provide flexibility and reliability by facilitating the integration of new power resources (such as renewable energy, wind, and solar energy), enabling corrective capabilities when failures occur, reducing carbon footprint, and reducing energy losses within the grid.

Smart grid is a system based on communication and information technology in generation, delivery, and consumption of energy power. It uses two-way flow of information to create an automated and widely distributed system that has new functionalities such as, real time control, operational efficiency, grid resilience, and better integration of renewable technology which will decrease carbon footprint. However, risks can still exist in smart grid. Any interruptions in power generation could disturb smart grid stability and could potentially have large socio-economic impacts. In addition, as valuable data are exchanged among smart grid systems, theft or alteration of this data could violate consumer privacy. Because of these weaknesses, smart grid has become the primary target of attackers [2], which attracted the attention of government, industry, and academia.

Several research papers have been published that provide an overview of the prevailing problems related to cyber security in smart grid infrastructure [3, 7]. In [3], Rawat et al. presented a study of the challenges present in smart grid security. They classified attacks based upon the type of the network, namely, home area network (HAN), neighborhood area network (NAN), and wide area network (WAN). In addition, they presented the impact of each attack on the information security: confidentiality, integrity, and availability (CIA). In [4], Shapsough et al. discussed security challenges in smart grid system, especially those related to connectivity, trust, customer privacy, and software vulnerabilities. The authors provided also an overview of the existing security solutions, particularly network security, data security, key management, network security protocols, and compliance checks. Another study, focusing on public networks, has been conducted by Liang et al. in [5]. The paper describes a protection framework of smart grid based on a public network. This framework was composed of three layers, main station, communication network, and terminals. In [6], Dari et al. discussed the security requirements and possible threats on smart grid. These threats were classified into three categories: people and policy, platform, and network threats. In [7], Wang et al. also classified attacks based on the CIA requirements, and they described several countermeasures, including network security, cryptographic, secure protocols, and secure architecture.

While these survey papers provide various classifications of attacks on smart grid, most of them are based upon confidentiality, integrity, or availability. However, blended and sophisticated attacks such as Stuxnet, Duqu, and Flame [8] can compromise all of the security parameters at the same time. Therefore, such attacks are usually excluded from these classification systems. Furthermore, countermeasures and security solutions were presented individually for each smart grid's component, and there is no global approach or process to combine all security mechanisms in order to ensure security for the entire system.

This paper provides a summary of the current status and future expectations of the smart grid cyber security. The remainder of this paper is organized as follows. First, we review cyber security objectives in smart grid. Next, we present a new classification system of cyber-attack based on a method used by hackers or penetration testers. This method allows one to better understand the process used by a hacker to compromise the smart grid security [9]. Then, we summarize and recommend a number of countermeasures. Some challenges and future directions are discussed in the last section.

## II. SMART GRID OVERVIEW

### A. Smart grid's features

The main benefits expected from the smart grid are increasing grid resilience and improving environmental performance. Resilience indicates the capability of a given entity to resist unexpected events and recover quickly thereafter [1]. Today, grid resilience as a feature has become nonnegotiable, especially when power interruptions can potentially impact the economy. Smart grid promises to provide flexibility and reliability by enabling additional dispersed power supply, facilitating the integration of new resources into the grid, and enabling corrective capabilities when failures occur. Moreover, smart grid systems are expected to enable electric vehicles as replacements for conventional vehicles, reducing energy used by customers and reducing energy losses within the grid [10].

### B. Smart grid's conceptual model

According to the national institute of standard and technology (NIST) [2], a smart grid is composed of seven logical domains: bulk generation, transmission, distribution, customer, markets, service provider, and operations, each of which include both actors and applications. Actors are programs, devices, and systems whereas applications are tasks performed by a one actor or more in each domain. Fig. 1 shows the conceptual model of smart grid and the interaction of actors from different domains via a secure channel.

Within the customer domain, the main actor is the end user. Generally, there are three types of customers: home, commercial/building, and industrial. In addition to consuming electricity, these actors may also generate, store, and manage the use of energy. This domain is electrically connected to the distribution domain and communicates with the distribution, operation, service provider, and market domains [2, 11].

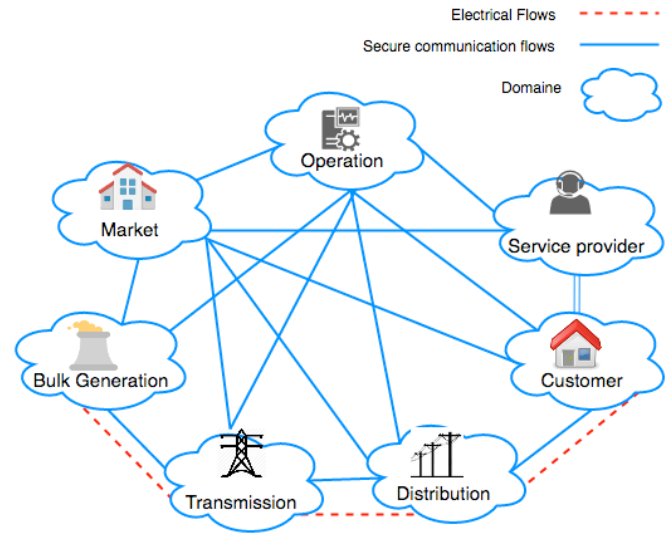

Fig. 1. Smart grid's conceptual model based on NIST.

In the market domain, actors are the operators and participants in the electricity markets. This domain maintains the balance between electrical supply and the demand. In order to match the production with demand, the market domain communicates with energy supply domains which include the bulk generation domain and distributed energy resources (DER) [2, 11]. The service provider domain includes the organizations that provide services to both electrical customers and utilities. These organizations manage services such as billing, customer account, and use of energy. The service provider interacts with the operation domain for situational awareness, system control and also communicates with customer and market domain to develop smart services such as enabling customer interaction with market and energy generation at home [2, 11]. The operations domain's actors are the managers of the movement of electricity. This domain maintains efficient and optimal operations in transmission and distribution. In transmission, it uses energy management systems (EMS), whereas in distribution it uses distribution management systems (DMS) [2, 11]. Actors in the bulk generation domain include generators of electricity in bulk quantities. Energy generation is the first step in the process of delivering electricity to the end user. Energy is generated using resources like oil, flowing water, coal, nuclear fission, and solar radiation. The bulk generation domain is electrically connected to the transmission domain and communicates through an interface with the market domain, transmission domain, and operations domain [2, 11]. In the transmission domain, generated electrical power is carried over long distances from generation domain to distribution domain through multiple substations. This domain may also store and generate electricity. The transmission network is monitored and controlled via a SCADA system, which is composed of a communication network, control devices, and monitoring devices [2, 11]. The distribution domain includes the distributors of electricity to and from the end user. The electrical distribution systems have different structures such as radial, looped, or meshed. In addition to distribution, this domain may also support energy generation and storage. This domain is connected to the transmission domain, customer domain, and the metering points for consumption [2, 11].

### C. Smart grid's systems

Smart grid is composed of several distributed and heterogeneous applications, including advanced metering infrastructure (AMI) [12], automation substation [13], demand response [13], supervisory control and data acquisition (SCADA), electrical vehicle (EV) [14], and home energy management (HEM) [13]. In this section we will discuss three critical and vulnerable applications in the smart grid: AMI, SCADA, and automation substation [1, 8, 12, 13, 15, 16, 17]. The other applications were discussed in detail in [12, 13].

Advanced metering infrastructure (AMI) is responsible for collecting, measuring and analyzing energy, water and gas usage. It allows two-way communication from the user to the utility. It is composed of three components: smart meter, AMI headend, and the communication network [18]. Smart meters are digital meters, consisting of microprocessors and a local memory, and they are responsible first for monitoring and collecting power usage of home appliances, and also for transmitting data in real time to the AMI headend in the utility side. An AMI headend is an AMI server consists of meter data management system (MDMS) [12]. The communication between the smart meters, the home appliances, and the AMI headend is defined through several communication protocols such as Z-wave and Zigbee [18].

Supervisory control and data acquisition (SCADA) is a system that measures, monitors and controls electrical power grid. It is typically used for large-scale environments. It consists of three elements: the remote terminal unit (RTU), master terminal unit (MTU), and human–machine interface (HMI) [19]. RTU is a device composed of three components: first one used for data acquisition, second one responsible for executing instructions coming for the MTU, and a third one designed for the communication. MTU is a device responsible for controlling the RTU. The HMI is a graphic interface for the SCADA system operator [19]. The communication within SCADA system is based on many industrial protocol including distributed network protocol v3.0 (DNP3) and IEC

61850 [20].

The substation is a key element in the power grid network. it performs several functions including receiving power from generating facility, regulating distribution, and limiting power surge [13]. It contains devices that regulate and distributes electrical energy such as a remote terminal unit (RTU), global positioning system (GPS), human–machine interface (HMI), and intelligent electronic devices (IEDs) [21]. The substation sends operation data to the SCADA for controlling the power system. Many operations are automated within the substation in order to increase the reliability of the power grid [1]. The communication between the automation substation and other devices in transmission and distribution is defined by the standard IEC 61850 [22].

### D. Smart grid's network protocols

Distributed and heterogeneous applications in smart grid require different communication protocols. Fig. 2 illustrates the smart grid network architecture and the protocol used within each network. In the home area network (HAN), home appliances uses ZigBee and Z-wave protocols [18]. In the neighborhood area network (NAN), devices are usually connected via IEEE 802.11, IEEE 802.15.4, or IEEE 802.16 standards [18]. In the wide area network (WAN) and in supervisory control and data acquisition (SCADA) applications, several industrial protocols are used specially distributed networking protocol 3.0 (DNP3) and modicon communication bus (ModBus) [20]. Within substation automation, protocol IEC 61850 is used [7]. In this section we will discuss two widely used yet vulnerable protocols in smart grid [22-25]: Modbus and DNP3. Bluetooth, Z-Wave, Zigbee, 6LoWPAN, WiMAX, IEC 61850 protocol, and power line communication are discussed in depth in [12, 14, 22].

Modicon communication bus (ModBus) is a 7 layer protocol of the model OSI; it was designed in 1979 to enable the process controller to communicate in real-time with computers. There are three types of Modbus: Modbus ASCII, Modbus RTU, and Modbus/TCP. In the first one, messages are coded in hexadecimal. Though it is slow, it is ideal for radio links and telephone communications. In the second one, the messages are coded in binary and it is used over RS232. In the third one, the masters and slaves uses IP addresses for communication [23]. In a SCADA system ModBus is a master-slave protocol responsible for exchanging instruction between one master, remote terminal unit (RTU) or master terminal unit (MTU), and several slave devices, such as sensors, drivers, and PLCs [23]. On one hand, Modbus is widely used in industrial architecture, because of its relative ease of use by communicating raw data without restriction of authentication, encryption, or any excessive overhead [26]. One the other hand, these features make it vulnerable and easily exploitable [23, 25].

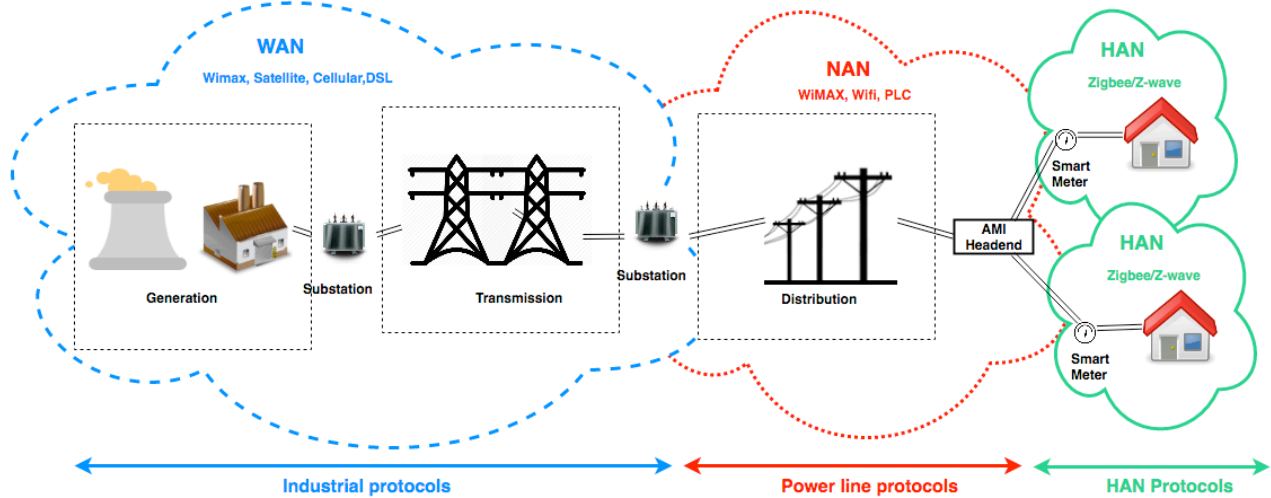

Fig. 2. Illustration of smart grid network architecture

Distributed network protocol v3.0 (DNP3) is another widely used communication protocol for critical infrastructure, more specifically in the electricity industry [24]. It was initiated in 1990 as a serial protocol to manage communication between "Master stations" and slave stations called "outstations' [26]. In electrical stations, DNP3 was used for connecting master stations, such as RTUs, with outstations, such as intelligent electrical devices (IEDs) [23]. In 1998, DNP3 was extended to work over IP network through encapsulation of TCP or UDP packets. DNP3 uses several standardized data formats and support timed-stamped (time-synchronized) data, making the data transmission reliable and efficient [26]. At first DNP3 did not provide any security mechanism such as encryption or authentication, but this problem was fixed with the secure version of DNP3 called DNP3 secure [7].

## III. SECURITY REQUIREMENTS OF SMART GRID

The National Institute of Standards and Technology (NIST) has defined three criteria required to maintain security of information in the smart grid and keep it protected, specifically confidentiality, integrity, and availability [10]. According to [27], accountability is another important security criterion. The description of each criterion is given below.

### A. Confidentiality

In general, confidentiality preserves authorized restrictions on information access and disclosure. In other words, the confidentiality criterion requires protecting both personal privacy and proprietary information from being accessed or disclosed by unauthorized entities, individuals, or processes. Once an unauthorized disclosure of information occurs, confidentiality is lost. For instance, information such as control of a meter, metering usage, and billing information that is sent between a customer and various entities must be confidential and protected; otherwise the customer's information could be manipulated, modified, or used for other malicious purposes [10].

### B. Availability

Availability is defined as ensuring timely and reliable access to and use of information. It is considered the most important security criterion in smart grid because the loss of availability means disruption of access to information in a smart grid [10]. For example, loss of availability can disturb the operation of the control system by blocking the information's flow through the network, and therefore denying the network's availability to control the system's operators.

### C. Integrity

Integrity in smart grid means protecting against improper modification or destruction of the information. A loss of integrity is an unauthorized alteration, modification, or destruction of data in undetected manner [10]. For example, power injection is a malicious attack launched by an adversary who intelligently modifies the measurements and relays them from the power injection meters and power flow to the state estimator. Both nonrepudiation and authenticity of information are required to maintain the integrity. Nonrepudiation means that individuals, entity or organization, are unable to perform a particular action and then deny it later; authenticity is the fact that data is originated from a legitimate source.

### D. Accountability

Accountability means ensuring tractability of the system and that every action performed by a person, device, or even a public authority is recordable so that no one can deny his/her action. This recordable information can be presented as an evidence within a court of law in order to determine the attacker [28]. An example of an accountability problem would be the monthly electricity bills of customers. Generally smart meters could determine the cost of electricity in real-time or day-to-day. However, if these meters are under attack this information is no longer reliable because they have been altered. As a result, the customer will have two different electric bills, one from the smart meter and the other from the utility [27].

## IV. SECURITY PROBLEMS AND COUNTERMEASURES IN SMART GRID

### A. Smart grid attacks

In general and as shown in Fig. 3, there are four steps used by malicious hackers to attack and get control over a system, namely reconnaissance, scanning, exploitation, and maintain access [9]. During the first step, reconnaissance, the attacker gathers and collects information about its target. In the second step, scanning, the attacker tries to identify the system's vulnerabilities. These activities aim to identify the opened ports and to discover the service running on each port along with its weaknesses. During the exploitation step, he/she tries to compromise and get a full control of the target. Once the attacker has an administrative access on the target, he/she proceeds to the final step which is, maintaining the access. This step is achieved by installing a stealthy and undetectable program; thus he/she can get back easily to the target system later.

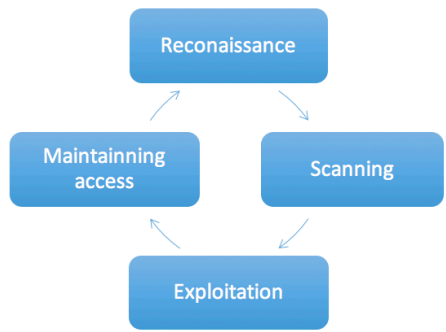

Fig. 3. Attacking cycle followed by hackers to get control over a system.

In smart grid, the same steps are followed by attackers to compromise the security's criteria [1]. During each step, they use different techniques to compromise a particular system in the grid. Thus, attacks can be classified based on these steps. Fig. 4 illustrates the types of attacks during each step. As one can see, numerous types of attacks can happen during the exploitation step. The malicious activities and attacks during each step described below.

#### 1) Reconnaissance

The first phase, reconnaissance, includes the attacks: social engineering and traffic analysis. Social engineering (SE), relies on social skills and human interaction rather than technical skills. An attacker uses communication and persuasion to win the trust of a legitimate user and get credential and confidential information such as passwords or PIN number to log on into a particular system. For examples, phishing [29] and password pilfering attack [30] are famous techniques used in SE. The traffic analysis attack is used to listen to the traffic and analyze it in order to determine the devices and the hosts connected to the network along with their IP addresses. Social engineering and traffic analysis compromise mainly the confidentiality of the information.

#### 2) Scanning

Scanning attack is the next step used to discover all the devices and the hosts alive on the network. There are four types of scans: IPs, ports, services, and vulnerabilities [9]. Generally, an attacker starts with an IPs scan to identify all the hosts connected in the network along with their IP addresses. Next, he or she goes deeper by scanning the ports in order to determine which port is open. This scan is executed on each discovered host on the network. The attacker then moves on to the service scan in order to find out the service or system running behind each opened port. For instance, if the port 102 is detected open on a particular system, the hacker could infer that this system is a substation automation control or messaging. If the port 4713 is open, the target system is a Phasor Measurement Unit (PMU) [1]. The final step, vulnerabilities scan, aims to identify the weaknesses and vulnerabilities related to each service on the target machine to exploit it afterward.

Modbus and DNP3 are two industrial protocols vulnerable to scanning attacks. Given that Modbus/TCP was designed for communication rather than security purpose, it can be compromised by an attack called Modbus network scanning [31]. This attack consists of sending a benign message to all devices connected in the network to gather information about these devices [31]. Modscan is a SCADA Modbus network scanner designed to detect open Modbus/TCP and identify device slave IDs along with their IP addresses [25]. Nicolas R. et al. have proposed an algorithm to scan the DNP3 protocol and discover hosts, specifically, the slaves, their DNP3 addresses, and their corresponding master [24]. As one can see, these attacks target mainly the confidentiality of the smart grid.

#### 3) Exploitation

The third step, exploitation, includes malicious activities that attempt to exploit the smart grid component's vulnerabilities and get the control over it. These attacks include viruses, worms, Trojan horses, denial of service (DOS) attacks, man-in-the-middle (MITM) attacks, replay attacks, jamming channels, popping the human machine interface (HMI), integrity violations, and privacy violations.

A virus is a program used to infect a specific device or a system in smart grid [1, 32]. A worm is self-replicating program. It uses the network to spread, to copy itself, and to infect other devices and systems [1, 32]. A Trojan horse is a program that appears to perform a legitimate task on the target system. However, it runs a malicious code in the background. An attacker uses this type of malware to upload a virus or worm on the target system [1, 32]. In June 2010, Roel Schouwenberg, a senior research at Kaspersky Lab, detected Stuxnet, the first worm targeting supervisory control and data acquisition (SCADA) systems [8]. This is regarded as the first cyber attack against a physical industrial control system.

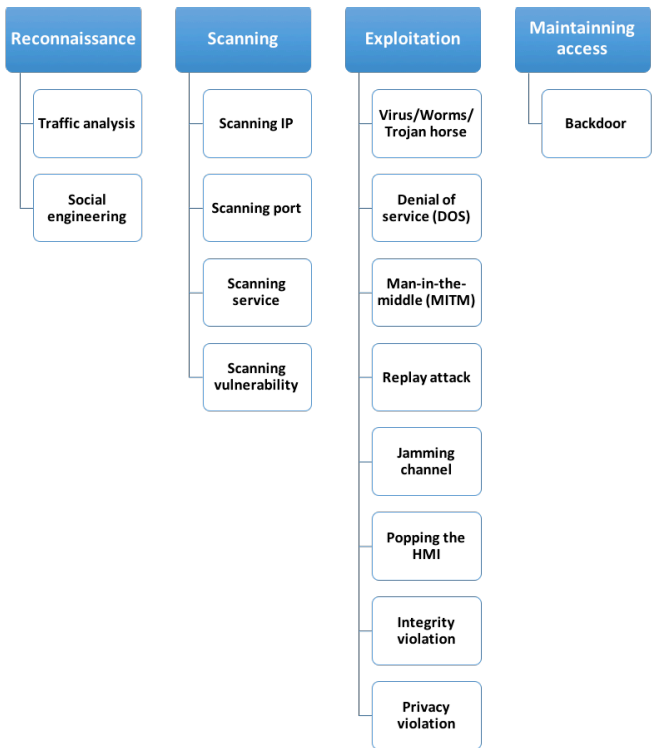

Fig. 4. Cyber attacks classification in smart grid based on the attacking cycle

Stuxnet, a worm of 500 Kilobytes, exploited many zero-days, which are software vulnerabilities that have not yet disclosed by the software owner. It infected at least 14 industrial sites based in Iran, including a uranium-enrichment plant. More than one year later, two more worms that targeted industrial control systems were discovered, Duqu and Flame. Unlike Stuxnet, Duqu was designed to gather and steal information about industrial control systems. Flame, on the other hand, was created to be used in cyber espionage in industrial networks. It has been found in Iran and other Middle East countries [1, 8, 33]. Viruses and worms can compromise availability, as Stuxnet did, confidentiality, as Duqu did, or a combination of the security's parameters.

In denial of service (DOS) attacks, several methods are used, particularly SYN attacks, buffer overflow, teardrop attacks, and smurf attacks [1, 32], puppet attack [15], time-delay-switch (TDS) [34], and time synchronization attack (TSA) [35]. A SYN attack exploits the three-way handshake (SYN, SYN-ACK, ACK) used to establish a TCP session. The attacker floods a target system with connection requests without responding to the replays, forcing the system to crash. The Modbus/TCP protocol is vulnerable to these attacks since it operates over TCP [26].

In buffer overflow, the attacker sends a huge amount of data to a specific system, thereby exhausting its resources. For example, the ping-of-death is considered as a buffer overflow attack as it exploits the internet control message protocol (ICMP) by sending more that 65K octets of data. It then makes the system crash.

In a teardrop attack, an attacker alters and modifies the length and the fragmentation offset fields in sequential IP packets. Once the target system receives these packets, it crashes because the instructions on how the fragments are offset within these packets are contradictory.

In smurf attack, the attacker targets not only a specific system, but it can saturate and congest the traffic of an entire network. It consists of three elements: the source site, the bounce site, and the target site. For source site, the adversary sends a spoofed packet to the broadcast address of the bounce site. These packets contain the IP address of the target system. Once the bounce site receives the forged packets, it broadcasts them to all hosts connected to the network and then causes these hosts to replay, saturating the target system [1, 32].

In puppet attack [15] targets the advanced metering infrastructure (AMI) network by exploiting a vulnerability in dynamic source routing (DSR) protocol and then exhausting the communication network bandwidth. Due to this attack, the packet delivery drops between 10% and 20%.

The time-delay-switch (TDS) [34] attack consists of introducing a delay in control system creating instability in the smart grid system.

The time synchronization (TSA) attack [35] targets mainly the timing information in smart grid. Because power grid operations such as fault detection and event location estimation depend highly on precise time information, and also most of the measurement devices in smart grid are equipped with global positioning system (GPS), attack such as TSA, which spoof the GPS information, could have a high impact on the system. DOS represents a significant threat to the smart grid system because communication and control messages in such a system are time critical [11], and a delay of few seconds could compromise the system availability.

The man-in-the-middle (MITM) attack is performed when an attacker inserts itself between two legitimate devices and listens, performs an injection, or intercepts the traffic between them. The attacker is connected to both devices and relays the traffic between them. These legitimate devices appear to communicate directly when in fact they are communicating via a third-device [1]. For example, an attacker could conduct a MITM, by placing himself on an Ethernet network to alter or misrepresent I/O values to the human machine interface (HMI) and programmable logic controllers (PLC). The MITM could also be used to intercept TCP/IP communication between the substation gateway and the transmission SCADA server [1]. Peter M. et al. have conducted several experiments, including the MITM attack, to compromise the integrity of the SCADA communication. They demonstrated the impact of such attack on the SCADA servers [17]. Ihab D. et al. [36] highlighted the vulnerability of the protocol DNP3 operating in SCADA, and they conducted an experiment of MITM attack with two scenarios showing that it is easy to intercept the messages exchanged between the master station and the outstations, modify the packet's content, and inject it into the network.

Intercept/alter attack is another type MITM attack. It attempts to intercept, alter, and modify data either transmitted across the network or stored in a particular device [33]. For example, in order to intercept a private communication in advanced metering infrastructure (AMI), an attacker uses electromagnetic/radio-frequency interception attack.

Eavesdropping attack is also another MITM attack's type, where the attacker intercepts private communications between two legitimate devices [33]. All these MITM attacks attempt to compromise the confidentiality, the integrity, and the accountability.

In replay attack, as the industrial control traffic is transmitted in plain text, an attacker could maliciously capture packets, inject a specific packet, and replay them to the legitimate destinations [1], compromising then the communication's integrity. Intelligent electronic device (IED),which is a device designed for controlling and communicating with the SCADA system [7], could be targeted by replay attacks so that false measurements are injected in a specific register [1]. Replay attack could also be used to alter the behavior the programmable logic controllers (PLC) [1]. In AMI, where an authentication scheme is used between smart meters, a replay attack involves a malicious host to intercept authentication packets sent from smart meter and re-sending them at a later point in time, expecting to authenticate and gain unauthorised entry into the network [37].

In the jamming channel attack, an adversary exploits the shared nature of the wireless network and sends a random or continuous flow of packets in order to keep the channel busy and then prevents legitimate devices from communicating and exchanging data [38]. Due to its time-critical nature, smart grid requires a highly available network to meet the quality of service requirements and such an attack can severely degrade its performance [38]. Keke G. et al. [16] proposed a jamming attack named maximum attacking strategy using spoofing and jamming (MAS-SJ) that targets mainly the cognitive radio network (CRN) in the wireless smart grid network (WSGN). Because WSGN is important for monitoring power grid in the smart grid with the PMU that plays a key component by providing time-synchronized data of power system operating states [39], attacks like MAS-SJ can disturb the operation of the system or even make it unavailable.

Popping the HMI is an attack that exploits a known device's vulnerability, especially device's software or OS vulnerabilities, and then installs a remote shell, allowing the attacker to connect remotely to the server from his computer to get unauthorized access in order to monitor and control the compromised system [1]. SCADA systems, substations, or any system running an operation system with a console interface is considered as a potential target of this attack. Even given the potential impact of such an attack, it does not require advanced networking skills or significant experience in security and industrial control system to perform. Since the devices' vulnerabilities documentation are publicly available, a hacker or the so-called script-kiddies may simply use open source tools such as Metasploit and meterpreter to launch such an attack and gain full control of the target system [1]. The availability, integrity, confidentiality, and accountability may be compromised based on the attacker's objective and motivation.

In the masquerade attack, a malicious person may pretend to be a legitimate user in order to gain access to a system or gain greater privileges to perform unauthorized actions. This attack could tamper with the programmable communicating thermostat (PCT) which is used to reduce electric power at a residential site. It compromises the availability, integrity, confidentiality, and accountability of the system [33].

Integrity violation attacks aim to violate the integrity and/or the accountability of the smart grid by altering intentionally or unintentionally the data stored in a given device in the network. For instance, a customer could perform this attack to alter the smart meter data in order to reduce his electricity bill. This attack could also be used to target remote terminal unit (RTU), so wrong data will be reported to the control center, resulting in an increased outage time. False data injection (FDI) [40] attack is a type of integrity violation. It aims to introduce arbitrary errors and corrupt some device's measurements, affecting the accuracy of the state estimate (SE). Since the SE is important for system monitoring to ensure reliable operation in the power system, and for the energy management system (EMS) to process a real-time data collected by the SCADA system, FDI attack could compromise the SE's integrity leading to the instability of the smart grid system [40]. A detailed study on the impact of the FDI attack on the power system stability was conducted by Adnan A. et al. in [41].

Privacy violation attack aims to violate privacy by collecting private information about customers [28]. For example, as smart meters collect electricity usage many times per hour, information about the user electricity's consumption could be obtained. Thus, if a meter does not show electricity usage for a period of time, that commonly indicates that the house is empty. This information could then be used to conduct a physical attack like burglary [28]. Depeng L. et al. showed that the demand response programs and smart meters generate high-resolution data about the customers' privacy. This data could be exploited leading to the loss of customer's information and disclosure of activity patterns [42].

### 4) Maintaining access

In the final step, maintaining access, the attacker uses a special type of attack to gain permanent access to the target, especially backdoors, viruses, and Trojan horses. A backdoor is an undetectable program, stealthy installed on the target to get back later easily and quickly [32]. If the attacker succeeds in embedding a backdoor into the servers of the control center of the SCADA, he or she can launch several attacks against the system which can cause a severe impact on the power system [43].

In IT network, security's parameters are classified based on their importance in the following order: confidentiality, integrity, accountability, and availability. Whereas in smart grid, they are classified: availability, integrity, accountability, and confidentiality [10, 28]. Thus, we can say that attacks which compromise the availability of the smart grid systems have a high severity, while those targeting confidentiality have a low severity. In addition to the level of severity, each attack has a level of likelihood to be performed. For instance, attacks such as Stuxnet and Duqu [8], has a high severity because they are able to vandalize the industrial control system and bypass all the security boundaries; but, they are complex and sophisticated. So, these viruses have high severity, but their likelihood to be performed is low. Another example is the HMI popping attack. It has a high severity and it does not require advanced networking skills or significant experience in security and industrial control system to perform it. Since the devices' vulnerabilities documentation are publicly available, a hacker or the so-called script-kiddies may simply use open source tools such as Metasploit and meterpreter to launch such an attack [1]. Therefore, this attack has high severity and it is very likely to be performed. Table II shows the likelihood of each attack to be performed and its associated level of severity.

## B. Smart grid countermeasures

A number of attack detection and countermeasure techniques are proposed in the literature to counter cyber attacks. For instance, Özçelik I. et al in [44] proposed a solution for distributed denial of service (DDos). In [38], Lu Z. et al. proposed a technique to detect jamming channel attacks. In [45] Rawat D. et al. proposed a technique to detect False detection injection (FDI) attacks. Though these security solutions contribute to the smart grid's security, they are insufficient to face sophisticated and blended attacks [1]. Moreover, Stuxnet [8] showed that strategy like "Defense in-depth" or "security by obscurity" [1] are no longer considered as valid solutions. We believe that security cannot be achieved through one specific solution, but by deploying several techniques incorporated into a global strategy. In this section, and as Fig. 5 shows, we propose a cyber security strategy composed of three phases: pre-attack, under attack, and post-attack. As follows, and for each phase, relevant published solutions in terms of security protocols, security technology, cryptography, and other cyber-attack countermeasures are described.

### 1) Pre-attack

During this first phase, pre-attack, various published solutions are recommended to enhance the smart grid's security and to be prepared for any potential attack. Security countermeasures commonly fall into three categories, namely network security, cryptography, and device security. We will discuss technologies and secure protocols such as IDS, SIEM, DLP and secure DNP3 [1, 18, 46] for the network security. Encryption, authentication, and key management [4, 7, 32, 47] for the data security. Finally, Host IDS, compliance checks, and diversity technique for the device security.

#### a) Network security

The network is the backbone of a smart grid. So, network security plays a significant role in securing the entire system. Using firewalls supplemented with other monitoring and inspection technologies is recommended [1] to secure the smart grid network. A firewall is intended to allow or deny network connections based on specific rules and policies. But an unknown or an advanced attack technique can easily bypass many firewall techniques. Therefore, firewalls should be associated with other security technologies such as intrusion detection system (IDS), security information and event management systems (SIEM), and network data loss prevention (DLP) [1, 18, 46]. IDS is a system developed for detecting malicious activity either on a network or on a specific host [32]. SIEMS are information management systems that collect and gather information such as operating system logs, application logs, and network flow from all devices in the network. Then the collected information will be analyzed and processed by a centralized server in order to detect any potential threat or a malicious activity in the network. Network DLP is a system responsible for preventing the loss or the theft of the data across the network [1].

| | | Severity of the attack | | |
|---|---|---|---|---|
| | | Low | Medium | High |
| **Likelihood of the attack to be performed** | High | • Traffic analysis [29, 30, 33] <br> • Privacy violation [28] | | • Virus, worms, Trojan horse [8] <br> • DOS [15, 35] <br> • Backdoor [43] |
| | Medium | • Social engineering [1, 29] <br> • Scanning [25, 28] | • MITM [17, 36] <br> • Replay attack [1, 34] | • Jamming channel [16, 38] <br> • Masquerade attack [33] <br> • Integrity violation [1, 41, 46] |
| | Low | | | Popping the HMI [1] |

TABLE I : LIKELIHOOD OF THE ATTACK TO BE PERFORMED AND ITS ASSOCIATED SEVERITY.

In addition to these security systems, secure network protocols such as IPsec, transport layer security (TLS), secure sockets layer (SSL), secure DNP3 can also be used to enhance security in the network. DNP3 is an industrial protocol widely used in smart gird [24]. Initially, DNP3 protocol came without any security mechanisms. In other words, messages are exchanged in plain text across the network and can be easily intercepted. In recent years, the increased number of cyber-attacks targeting industrial and power system has attracted the attention of a number of researchers in both industry and academia. Consequently, a secured variation of DNP3 protocols has been released named secure DNP3.

This secured version added a secure layer for encryption and authentication between the TCP/IP and application layer. Using such a protocol, several attacks can be avoided, for example, authentication mechanism can protect against MITM attack, whereas encryption decreases eavesdropping and replay attacks. Secure DNP3 is discussed further in [7].

### b) Cryptography for data security

Encryption mechanisms aim to ensure data's confidentiality, integrity, and nonrepudiation. There are two types of key encryptions: symmetric and asymmetric. In symmetric key encryption, or single-key encryption, one key is used to encrypt and to decrypt data. The most used algorithms employing symmetric encryption are advanced encryption standard (AES) and data encryption standard (DES). Asymmetric key encryption, on the other hand, uses two keys to encrypt and decrypt data: private key and public key. RSA (Rivest, Shamir and Adleman) is a widely used asymmetric algorithm [32]. In smart grid, various components with different computational capabilities co-exist. Therefore, both symmetric and asymmetric key encryption can be used, and the selection depends on several factors, including data criticality, time constraints, and computational resources [4].

Authentication is defined as the act of verifying that an object's identity is valid, such as the use of a password [32]. An object could be a user, a smart device, or any component connected to the smart grid network. Multicast authentication is a particular type of authentication and its applications are widely used in smart grid [47]. In [4], Shapsough et al. proposed three methods to achieve authentication for multicast applications: secret-info asymmetry, time asymmetry, and hybrid asymmetry.

Key management is a crucial approach for encryption and authentication. Public key management (PKI), or shared secret key management, can be used to ensure authenticity for communication across networks. In PKI infrastructure, the identities of two parties is verified by a certificate delivered from a third party called the certificate authority (CA). This mechanism is done before establishing any connection between the two parties. In shared secret key management, four steps are used to maintain communication security: key generation, key distribution, key storage, and key update [4]. Due to the distributed nature of smart grid, some specific requirements should be considered to design a cryptography

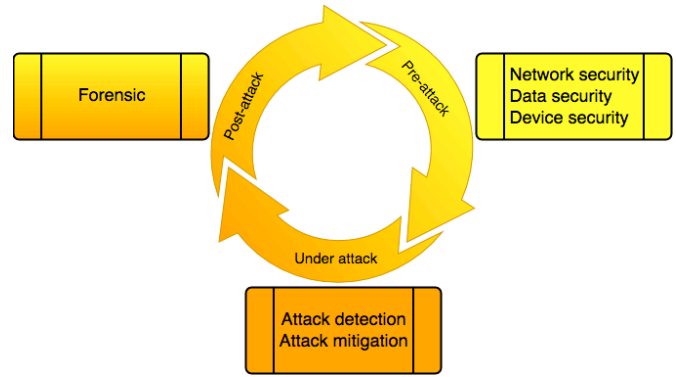

Fig. 5. Cyber security strategy for smart grid

key management, W. Wang et al. in [7] present several basic yet relevant requirements of the key management scheme, particularly efficiency, evolve-ability, scalability, and secure management. In addition, several key management frameworks have been proposed specifically for the power system: single-key, key establishment scheme for SCADA systems (SKE), key management architecture for SCADA systems (SKMA), advanced key management architecture for SCADA systems (ASKMA), ASKMA+, and scalable method of cryptographic key management (SMOCK) to name a few. The choice of a framework relies on different criteria, including scalability, computational resource capability, and support for multicast. The authors conducted a comparison between the key management schemes listed above. The comparison was based on scalability, support for multicast, robust to key compromise, and power system application. ASKMA+ and SMOCK show interesting results. ASKMA+ is an efficient key management scheme and it supports multicast, but it still suffers from scalability. SMOCK, on the other hand, shows good scalability; however, it has some weaknesses such as no support for multicast and low computational efficiency.

### c) Device security

Device protection is the third crucial element in the supply chain of smart grid security. Many research papers and recommendation reports have been published contributing to security assurance for endpoints. In [1], several security technologies have been recommended, particularly, host IDS, anti-virus, and host data loss prevention (DLP). Additionally, Kammerstetter et al. [48] recommended using an automated security compliance check. Such a tool performs a check against all smart grid components to verify that each device's configuration is up to date, especially the device's firmware and the current configuration file. As the smart grid components are highly connected and a weakness in one component can expose the entire system to risk, a compliance check is a crucial tool. In [49], Mclaughlin et al. proposed a diversity technique in smart meter's firmware to limit a large-scale attack. Using such a technique, an attacker can exploit a vulnerability of one device's firmware, but he or she cannot exploit the same vulnerability on other devices.

*2) Under attack*

This step is divided into two tasks: attack detection and attack mitigation. Several approaches and technologies can be used during each task, to detect the malicious activity, and then deploy the appropriate countermeasures.

During the attack detection, all the deployed security technologies are recommended, including SIEMS, DLP, and IDS [1, 18, 46]. But, some of these solutions have a number of limitations and need improvements, particularly IDS. IDS is a widely used security system in IT network, and it is also used in smart grid network; but, it has many performance limitations specially reporting high rate of false positive. Thus, many research papers were published to improve the IDS performance in the smart context [46, 50, 51]. Y. Kwon et al. [50] proposed an approach based on IDS for the IEC 61850 protocol. They used both statistical analysis and specification-based metrics. The experimental results showed that their approach could detect anomalies in large networks with low false positive. In [51], U. K. Premaratne et al.  proposed an IDS IEC61850 automated substations. This rule-based IDS was developed based on collecting data from simulated attacks on an IED. The result of the experiment showed that the IEC61850 IDS was capable of detecting many attacks such as a DOS attack, a password cracking attack, and an ARP packet sniffer attack. Faisal M. et al. [18] proposed an architecture of IDS in AMI based on stream mining algorithms. They conducted an experiment to compare the seven existing state-of-the-art data stream mining algorithms: Accuracy Updated Ensemble, Active Classifier, Leveraging Bagging, Limited Attribute Classifier, Bagging using ADWIN, Bagging using Adaptive Size Hoeffding Tree, and Single Classifier Drift. Their comparison was based on several metrics including execution time, detection accuracy, and memory consumption. For the assessment they used an original version of the KDD Cup 1999 [52] and an improved version of this data set. The results showed that some algorithms do not require an advanced computational resources, so they are suitable for IDS in some devices such as smart meters. Other algorithms have a high accuracy and they require more computational resources; these algorithms can be used for the IDS in a data concentrator or in an AMI headends [18]. Zhang Y. et al. [46] proposed a distributed intrusion detection system for smart grids (SGDIDS) based on an intelligent model. This model can be used in every level of the smart grid: home area network (HAN) , neighborhood area network (NAN), and wide area network (WAN). The proposed IDS was based on data mining algorithms: support vector machine (SVM) and artificial immune system (AIS). To evaluate the efficiency of their solution, they used a simplified and an improved version of the KDD cup 1999 called the NSL-KDD [53]. The combination of two classifiers SVM and AIS have produced satisfactory results in terms of detection malicious traffic [46].

Once the attack are detected, mitigation can be executed using the following methods. S. Shapsough et al. [4] surveyed and summarized several methods used to mitigate the DOS attack, especially pushback and reconfiguration methods. In pushback, the router is configured to block all the traffic coming from the attacker's IP address. In the reconfiguration method, the network topology is changed to isolate the attacker. For jamming attacks, Lu et al. [38] discussed anti-jamming schemes such as frequency hopping spectrum spread (FHSS) and direct sequence spectrum spread (DSSS) to mitigate attacks. Other mitigation techniques for buffer overflow, man-in-the-middle, CPU exhausting, and replay attack, distributed denial of service (DDos), and false data injection (FDI) were discussed in detail in [3, 27, 44].

*3) Post-attack*

When an attack is not detected, such as in the case of Stuxnet [8], the post-attack period is an important step. First, it is critical to identify the entity involved in the attack. Then, the IDS signature, anti-virus database and security policies must be kept up to date by learning from attacks and to protect the smart grid against future similar attacks. Forensic analysis is the primary technique used during the post-attack. Smart grid forensic studies collect, analyze, and intercept digital data in order to identify the entity involved in the event. They are also useful to determine and address cyber and physical vulnerabilities of the smart grid in order to anticipate potential attacks. In addition, forensic analysis in smart grid plays an important role in the investigation of cyber-crimes such as hacking, viruses, digital espionage, cyber terrorism, manipulating the operation of the smart grid, violating the consumer's privacy, and stealing valuable information including intellectual property and state secrets [54].

Table II shows a summary of the cyber attacks in smart grid based upon the fours steps: reconnaissance, scanning, exploitation, and maintaining access. Each step includes attacks' categories, attacks' examples, the compromised component in the smart grid by each attack, the impact of each attack, and the appropriate countermeasures. As we can see, the most attacks can be avoided by using secure network protocols such as secure DNP3, and also by enabling encryption and authentication mechanisms.

## V. CHALLENGES AND FUTURE DIRECTION

In heterogeneous systems such as smart grid, different devices coexist and communicate through various network protocols. This heterogeneity represents a great challenge and a potential threat for the smart grid security. The communication between devices requires aggregation of data and translation between protocols. However, this aggregation can enable accidental breaches and vulnerabilities simply because a feature in one protocol could not be translated properly into another [4].

| Attacking cycle step | Attack category | Attack example | Compromised application/protocol in smart grid. | Compromised security's parameter | Possible countermeasures |
|---|---|---|---|---|---|
| Reconnaissance | Traffic analysis | [33] | Modbus protocol, DNP3 protocol | Confidentiality | Secure DNP3, PKI (SKMA, SMOCK), TLS, SSL, Encryption, Authentication[1, 7] |
| | Social engineering | Phishing [29] | | | |
| | | Password pilfering [30] | | | |
| Scanning | Scanning IP, Port, Service, Vulnerabilities | Modbus network scanning [25] | Modbus Protocol | Confidentiality | IDS, SIEM, Automated[1] security compliance checks [48] |
| | | DNP3 network scanning [24] | DNP3 Protocol | | |
| Exploitation | Virus, worms, Trojan horse | Stuxnet [8] | SCADA PMU, Control device | Confidentiality Integrity Availability Accountability | DLP , IDS , SIEM, Anti-virus [1], Diversity technique[49] |
| | | Duqu [8] | SCADA | | |
| | Denial of service (DOS) | Puppet attack [15] | AMI | Availability | SIEM, IDS [1], flow entropy, signal strength, sensing time measurement, transmission failure count, pushback, reconfiguration methods [4, 44] |
| | | TDS [34] | Instability of smart grid systems | | |
| | | TSA [35] | PMU, smart grid equipment's GPS | | |
| | Man-in-the-middle (MITM) | eavesdropping attack [1, 33] | HMI, PLC | Confidentiality Integrity | Secure DNP3, PKI (SKMA, SMOCK) [7], TLS, SSL, encryption, authentication [1] |
| | | [17] | SCADA | | |
| | | [36] | DNP3, SCADA | | |
| | | Intercept/alter [33] | AMI | | |
| | Replay attack | [1] | IED, SCADA, PLC | Confidentiality Integrity | Secure DNP3, TLS, SSL, encryption, authentication[1] PKI (SKMA, SMOCK) [7], |
| | | [33] | Authentication scheme in AMI | | |
| | Jamming channel | [38] | PMU | Availability | JADE, anti-jamming (FHSS, DSSS) [38] |
| | | MAS-SJ [16] | CRN in WSGN | | |
| | Popping the HMI | [1] | SCADA, EMS, substations. | Confidentiality Integrity Availability Accountability | DLP, IDS , SIEM , Anti-virus [1], automated security compliance checks [48] |
| | Masquerade attack | [33] | PLC | Confidentiality Integrity Availability Accountability | DLP, IDS, Secure DNP3, SIEM, TLS, SSL, encryption, authentication [1], PKI (SKMA, SMOCK)[7] |
| | Integrity violation | [1] | Smart meter, RTU | Integrity Availability | DLP, IDS ,SIEM, Secure DNP3, TLS, SSL, encryption, authentication [1], PKI (SKMA, SMOCK) [7, 45] |
| | | FDI [40], [41] | EMS, SCADA, AMI | | |
| | Privacy violation | [28], [42] | Demand Response program, Smart meters. | Confidentiality | Secure DNP3, PKI (SKMA, SMOCK)[7], TLS, SSL, encryption, authentication [1] |
| Maintaining access | Backdoor | [43] | SCADA | Confidentiality Integrity Availability Accountability | IDS, SIEM,Anti-virus [1], Diversity technique[49] |

TABLE II : CYBER ATTACKS IN SMART GRID, THEIR IMPACTS AND COUNTERMEASURE

Furthermore, the majority of industrial network protocols used in smart grid such as, DNP3, ICCP, Modbus, and Profibus, were designed for connectivity but not for security purposes. Thus, these protocols not only cannot ensure a secure communication channel, but they may also be used as an attack surface. Though there are some secure version of many industrial protocols, such as secure DNP3. However, the problem with this new version is its incompatibility with legacy installations [26].

In addition to network protocols, operating systems and physical equipment in smart grid may be vulnerable and expose the system to a wide variety of attacks. Since operating systems are designed for control in automation control components, they lack security features. Moreover, most of the physical devices are obsolete whereas others have insufficient memory space and limited computational capacity, so they cannot support advanced security mechanisms. For instance, smart meters have limited memory and computational resources because they are designed for lower power consumption, so they cannot support some important security mechanisms such as proper random number generators and cryptographic accelerators [55]. Although these components have less impact on the smart grid operation, if they are compromised, they represent a potential vector to compromise the whole system.

Security solutions such as IDS, firewalls, and encryption methods play a significant role in securing the conventional networks . However, these mechanisms have many limitations and they are inappropriate for a distributed environment with different application requirements such as latency and bandwidth [13]. In addition, these solutions are unable to counter the newest types of cyber-attacks. Since cyber-attacks are becoming more blended, sophisticated, and complex, they are able to target at the same time multiple layers of a communication system. For example, as previously mentioned, Stuxnet [8] was able to vandalize an industrial control system by bypassing all the security boundaries, demonstrating that the security solutions deployed in those scenarios are unable to detect such an effective virus. Furthermore, because there are several logical domains in smart grid (generation, transmission, distribution, markets, customer, and service provider), security requirements necessarily differ from one domain to another. For instance, in the generation domain denial of service (DOS) attacks need fast detection, which is not the case for market domain, customer domain, or service provider domain. In addition, the transmission domain requires delay-efficient key management, whereas the market domain requires large scale key management [7].

Therefore, rather than applying a simple security approach or deploying a specific security technology, we believe that smart grid cyber-attacks may be mitigated more effectively by combining several security mechanisms through a cyber security strategy. Such a strategy have several benefits, including, addressing the system's vulnerabilities, detecting a number of cyber-attacks, deploying the appropriate countermeasures, and identifying the involved entity.

## VI. CONCLUSION

Smart grid is a system composed of distributed and heterogeneous components to intelligently deliver the electricity and easily integrate the renewable technologies. However, this critical system suffers from a number of security weaknesses.   In this paper, we provide a comprehensive overview of cyber-security in smart grid and investigate in depth the main cyber-attacks threating its infrastructure, its network protocols, and its applications. In addition, we propose a strategy composed of several tools and mechanisms designed to address potential components' vulnerabilities, detect malicious activities, enhance communication security in the network, and protect the customer's privacy.